\documentclass[
journal=inoraj,
manuscript=article,
layout=twocolumn
]{achemso}

\usepackage[T1]{fontenc} 
\usepackage{bm}
\usepackage{siunitx}
\usepackage[version=4]{mhchem}
\usepackage{here}

\newcommand{\tit}[1]{\textit{#1}}
\newcommand{\trm}[1]{\textrm{#1}}
\newcommand{\mrm}[1]{\mathrm{#1}}

\author{Hiroki Ninomiya}
\email{hiroki.ninomiya@aist.go.jp}
\phone{+81 (0)29 8612312}
\fax{+81 (0)29 8615569}

\author{Kunihiko Oka}

\author{Izumi Hase}

\affiliation[AIST]
{National Institute of Advanced Industrial Science and Technology (AIST), Tsukuba, Ibaraki 305-8568, Japan}

\author{Kenji Kawashima}
\affiliation[IMRA]
{IMRA Material R\&D Co., Ltd., Kariya, Aichi 448-0032, Japan}
\alsoaffiliation[AIST]
{National Institute of Advanced Industrial Science and Technology (AIST), Tsukuba, Ibaraki 305-8568, Japan}

\author{Hiroshi Fujihisa}

\author{Yoshito Gotoh}

\author{Shigeyuki Ishida}

\author{Hiraku Ogino}

\author{Akira Iyo}

\author{Yoshiyuki Yoshida}

\author{Hiroshi Eisaki}

\affiliation[AIST]
{National Institute of Advanced Industrial Science and Technology (AIST), Tsukuba, Ibaraki 305-8568, Japan}

\title[An \textsf{achemso} demo]{
	Superconductivity in a Scandium Borocarbide with a Layered Crystal Structure
  }

\begin{document}



\begin{abstract}

The discovery of nearly room-temperature superconductivity in superhydrides has motivated further materials research for conventional superconductors. 
To realize the moderately high critical temperature $(T_\mathrm{c})$ in materials containing light elements, we explored new superconducting phases in a scandium borocarbide system. 
Here, we report the observation of superconductivity in a new ternary Sc--B--C compound. 
The crystal structure, which was determined through a Rietveld analysis, belongs to tetragonal space group $P4/ncc$. 
By complementarily using the density functional theory calculations, a chemical formula of the compound was found to be expressed as Sc$_{20}$C$_{8-x}$B$_x$C$_{20}$($x=1$ or 2). 
Interestingly, a small amount of B is essential to stabilize the present structure. 
Our experiments revealed the typical type-II superconductivity at $T_\mathrm{c}=7.7$ K. 
Additionally, we calculated the density of states within a first-principles approach and found that the contribution of the Sc-$3d$ orbital was mainly responsible for the superconductivity. 

\end{abstract}

\section{Introduction}

Exploring new superconductors with a higher critical temperature ($T_\mrm{c}$) is still one of the main topics in the fields of materials science and condensed matter physics. 
In the case of conventional phonon-mediated superconductors, the pairing mechanism of which is explained by the Bardeen-Cooper-Schrieffer (BCS) model,\cite{BCS-theory} it is widely accepted that $T_\mrm{c}$ can be enhanced by large density of states (DOS) at the Fermi energy, strong electron-phonon coupling, and high-frequency phonons. 
To realize a higher $T_\mrm{c}$, materials consisting of nonmagnetic metals and light elements are essential. 
In particular, there are several reports on the superconductivity in alkali, alkali-earth, and transition-metal borides and carbides. 
Typical examples and their critical temperatures (in parentheses) are summarized in Table \ref{Tc-summary}. 
The covalent characters of B and C, with high Debye frequencies owing to their light mass, result in the moderately high temperature superconductivity.\cite{MgB2_CaC6_covalent} 

The well-known magnesium diboride, \ce{MgB2}, crystallized in an \ce{AlB2}-type structure, which includes a honeycomb layer consisting of B, exhibits the highest $T_\mrm{c}$ of \SI{39}{K} in metals\cite{MgB2-Nagamatsu}. 
Electronic-structure studies experimentally as well as theoretically proposed that superconductivity in \ce{MgB2} is attributed to the $\sigma$- and $\pi$-bands derived from the B-$sp^2$ orbitals\cite{MgB2_BandStructure,MgB2-MultiGap}. 
On the other hand, the yttrium sesquicarbide of \ce{Y2C3} has a body-centered cubic unit cell\cite{Y2C3_original}, and its $T_\mrm{c}$, strongly depending on sintering conditions such as pressure and temperature, reaches \SI{18}{K}\cite{Y2C3-18K_JPSJ}. 
Unlike the case of \ce{MgB2}, such a relatively high $T_\mrm{c}$ is mainly due to the Y-$4d$ orbitals rather than the contribution from the C-$2p$ orbitals\cite{Y2C3_Band_JPSJ2007}. 
Moreover, in recent studies on hydrides at megabar pressures, nearly room-temperature superconductivity has been realized in \ce{H3S}($T_\mrm{c}=\SI{203}{K}$)\cite{H2S_203K_nature} and \ce{LaH_{10$\pm$$x$}}($T_\mrm{c}>\SI{260}{K}$)\cite{LaH10_260K_PRL}. 
Triggered by the discovery of such materials, attention has been focused on the search for conventional superconductivity including light elements. 

\begin{table}[!t]
\centering
\caption{
	Known Carbide and Boride Superconductors Combined with Alkali, Alkali-Earth, and Rare-Earth Metals. \textsuperscript{\emph{a}}
	}
\label{Tc-summary}
\begin{tabular}[width=\linewidth]{ccc}
\hline
metals & borides & carbides \\ \hline
Li  & \ce{Li_{1.7}B12}(\SI{7.1}{K})\cite{Li1.7B12_PRB} & \ce{LiC2}(\SI{1.9}{K})\cite{LiC2_Tc=1.9K}\\
Na & ---  & \ce{NaC2}(\SI{5.0}{K})\cite{NaC2_Tc=5K}\\
K &  ---  & \ce{K3C60}(\SI{19.3}{K})\cite{K3C60_Tc=19.3K} \\
Rb & --- & \ce{Rb_{$x$}C60}(\SI{28}{K})\cite{RbxC60_Tc=28K}  \\
Cs & --- & \ce{Cs3C60}(\SI{38}{K})\cite{Cs3C60} \\
Be & \ce{BeB_{2.75}}(\SI{0.79}{K})\cite{BeB2.75_Tc=0.79K} & --- \\
Mg &  \ce{MgB2}(\SI{39}{K})\cite{MgB2-Nagamatsu} & ---   \\
Ca  &  ---  & \ce{CaC6}(\SI{11.5}{K})\cite{CaC6_11.5K_PRL2005} \\
Sr &  ---  & \ce{SrC6}(\SI{1.65}{K})\cite{SrC6_PRL2007} \\
Ba  &  --- & \ce{Ba6C60}(\SI{7}{K})\cite{Ba6C60_Tc=7K}\\
Sc &  --- &  --- \\
Y &  \ce{YB6}(\SI{7.5}{K})\cite{YB6} & \ce{Y2C3}(\SI{18}{K})\cite{Y2C3-18K_JPSJ}\\
La & --- & \ce{La2C3}(\SI{13.5}{K})\cite{La2C3_1969} \\ \hline
\end{tabular}%

\textsuperscript{\emph{a}} The $T_\mrm{c}$ of these materials are in parentheses. 
\end{table}

In the present study, we explored new superconducting phases in the Sc--B--C system. 
Although a moderately high-$T_\mrm{c}$ superconductor is expected in combination with the comparatively small ionic radius of Sc and the potentially high Debye frequencies originating from B and C, there are no reports on the observation of superconductivity thus far, as listed in Table \ref{Tc-summary}. 
The phase relation in this system has been investigated by means of borothermal and carbothermal reductions, solid state reactions, and an arc-melting method\cite{Sc-B-C_system,Sc3B0.75C3-PhysProp,Sc2BC2_band}. 
Although there are few reports on the physical properties of Sc--B--C compounds, a metal-to-insulator transition associated with the disorder-driven Anderson localization is suggested in \ce{Sc3B_{0.75}C3}\cite{Sc3B0.75C3-PhysProp}. 

In this paper, we report on the successful synthesis, crystal structure, and superconductivity of the new Sc-based borocarbide. 
Utilizing the arc-melting technique, we can easily handle the higher melting-point materials, and extract metastable phases through the quenching effect. 
We found that the compound has a tetragonal layered structure with a composition of \ce{Sc20C_{8-$x$}B_{$x$}C20}$(x=1\ \trm{or}\ 2)$.
Magnetic, transport, and thermal measurements revealed the fully gapped superconductivity at $T_\mrm{c}=\SI{7.7}{K}$, where the electron-phonon coupling strength is moderately enhanced. 
Additionally, the DOS curves were calculated based on first principles. 

\section{Experimental Procedures}
\subsection{Sample Preparation and Crystal-Structure Analysis}
Polycrystalline samples were synthesized using the arc-melting method in an inert gas environment. 
As starting materials, we used powders of graphite (C) and amorphous boron (B), and ingots of scandium (Sc) prepared in a specified molar ratio. 
B and C were preliminarily mixed in a zirconia mortar and then pelletized so that the Sc ingot was sandwiched between the mixture. 
These procedures were performed in a dry nitrogen-filled glovebox to prevent the oxidation of the Sc metals. 
The prepared pellet was arc-melted on a water-cooled copper hearth in an argon atmosphere. 
To ensure chemical homogeneity, we turned over and remelted the button at least thrice. 
The resultant sample was metallic silver in color. 
The powder X-ray diffraction (PXRD) pattern was measured using Cu K$\alpha$ radiation. 
The intensity data were collected at approximately \SI{293}{K} using a diffractometer (Ultima-IV, Rigaku) equipped with a one-dimensional high-speed detector (D/teX Ultra, Rigaku). 
To refine the crystallographic parameters, we conducted Rietveld analyses using BIOVIA Materials Studio Reflex software(version 2018 R2)\cite{BIOVIA}. 

\subsection{Physical Property Measurements}
The magnetic susceptibility was measured using a commercial SQUID magnetometer (MPMS-XL7, Quantum Design) under a magnetic field of $H=\SI{10}{Oe}$. 
The data were collected on warming after zero-field cooling (ZFC process) and then cooling under the field (FC process). 
Additionally, we measured the isothermal magnetization in the field, ranging from $-30$ to \SI{30}{kOe}. 
Electrical-resistivity measurements down to $T=\SI{2}{K}$ were performed by the conventional four-probe technique at several fields between $H=0$ and \SI{90}{kOe} using a physical property measurement system (PPMS, Quantum Design). 
Furthermore, using the PPMS, we measured the specific heat below \SI{20}{K} using the thermal-relaxation method under zero field and \SI{90}{kOe}. 

\subsection{Electronic-Structure Calculation}
We conducted first-principles calculations using the full-potential linearized augmented plane-wave method (FLAPW) on the basis of density functional theory (DFT). 
The computer program WIEN2k\cite{WIEN2k} was used to implement the whole calculation. 
The exchange-correlation potential was approximated by the general-gradient approximation (GGA)\cite{GGA}. 
We set the $RK_\mrm{max}$ parameter as 7.0, where $R$ and $K_\mrm{max}$ represent the smallest muffin-tin radius in the unit cell and a cutoff parameter for the plane wave, respectively. 
The spin-orbit interaction and magnetism were not considered in this calculation. 

\section{Results and Discussion}

\subsection{Synthesis and Crystal-Structure Determination}

\begin{figure}[!t]
	\includegraphics[width=8cm,bb=0 0 320 339]{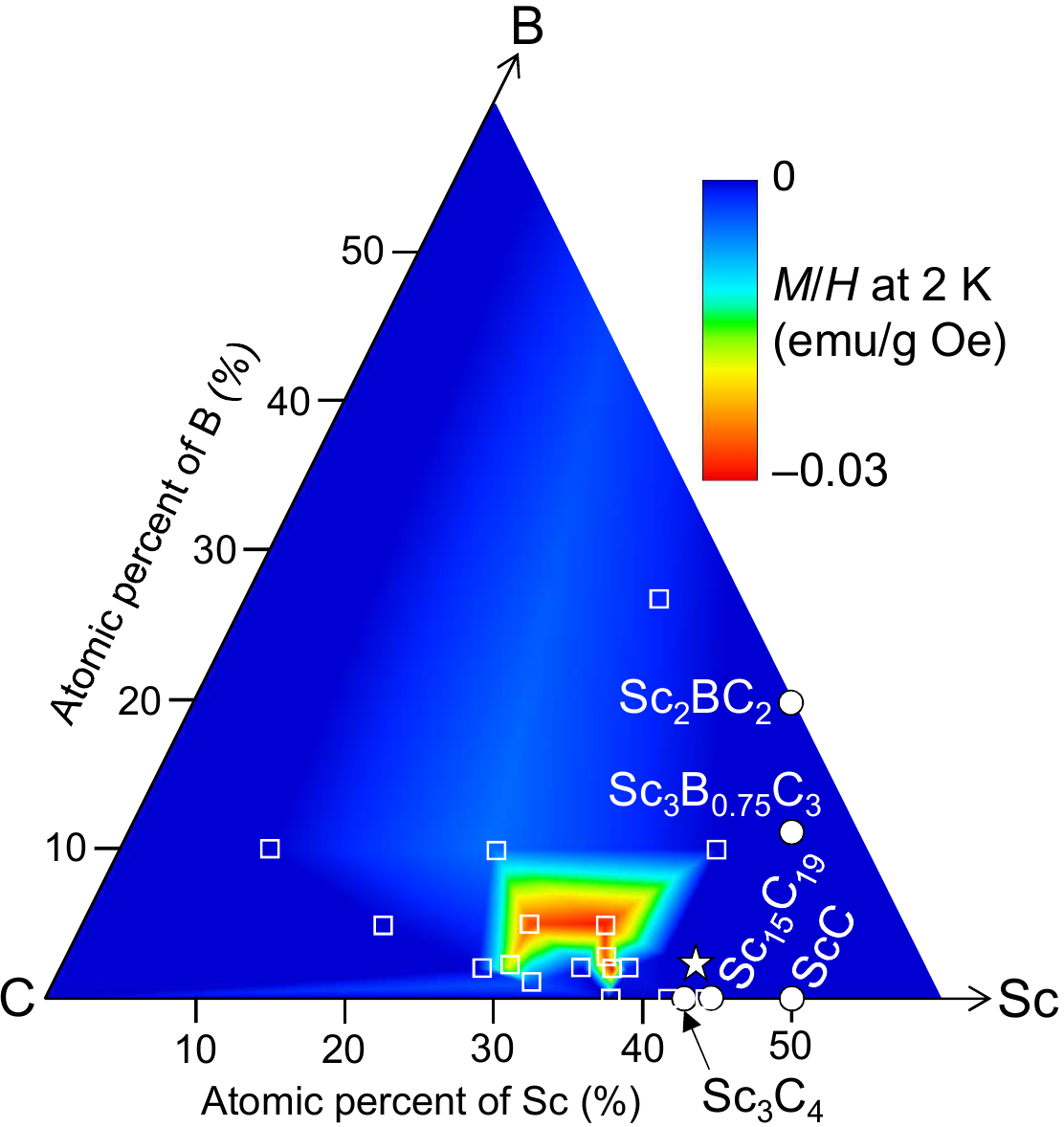}
	\caption{
	Magnified ternary phase diagram of the Sc--B--C system near the C-rich region. 
	Closed circles and open squares represent the already reported compounds and the tested compositions, respectively. 
	Contour plots indicate the magnitude of magnetic susceptibility $(M/H)$ defined as the magnetization divided by the field at a temperature of \SI{2}{K}. 
	Warmer or colder colors indicate that the sample prepared in its composition exhibited a larger or smaller diamagnetic signal, respectively. 
	The determined superconducting phase is represented by the star. 
	}
	\label{contour}
\end{figure}

We attempted to synthesize the ternary Sc--B--C system under various conditions of the starting composition, and then we examined the superconductivity by measuring the magnetic susceptibility down to \SI{2}{K}. 
The magnitude of the diamagnetic signals for each sample is depicted as the contour plot in Figure \ref{contour}. 
When the samples were prepared in the molar ratio within a hot spot, we found that a large Meissner signal was observed below approximately \SI{8}{K}. 
When we prepared the B-excess and/or B-free samples with the ratio of Sc to C set to 30--\SI{40}{\%}, none of the samples exhibited superconductivity down to \SI{2}{K}. 
This indicates that a small amount of B is necessary for the phase formation; however, it is difficult to evaluate the content of B using compositional analyses, such as energy-dispersive X-ray spectrometry. 
The chemical formula of the present superconducting phase, as is discussed below, is indicated by the star in Figure \ref{contour}. 
The discrepancy between the nominal and actual compositions may be attributed to the evaporation of C during arc melting. 

\begin{figure}[!t]
	\includegraphics[width=8.5cm,bb=0 0 494 419]{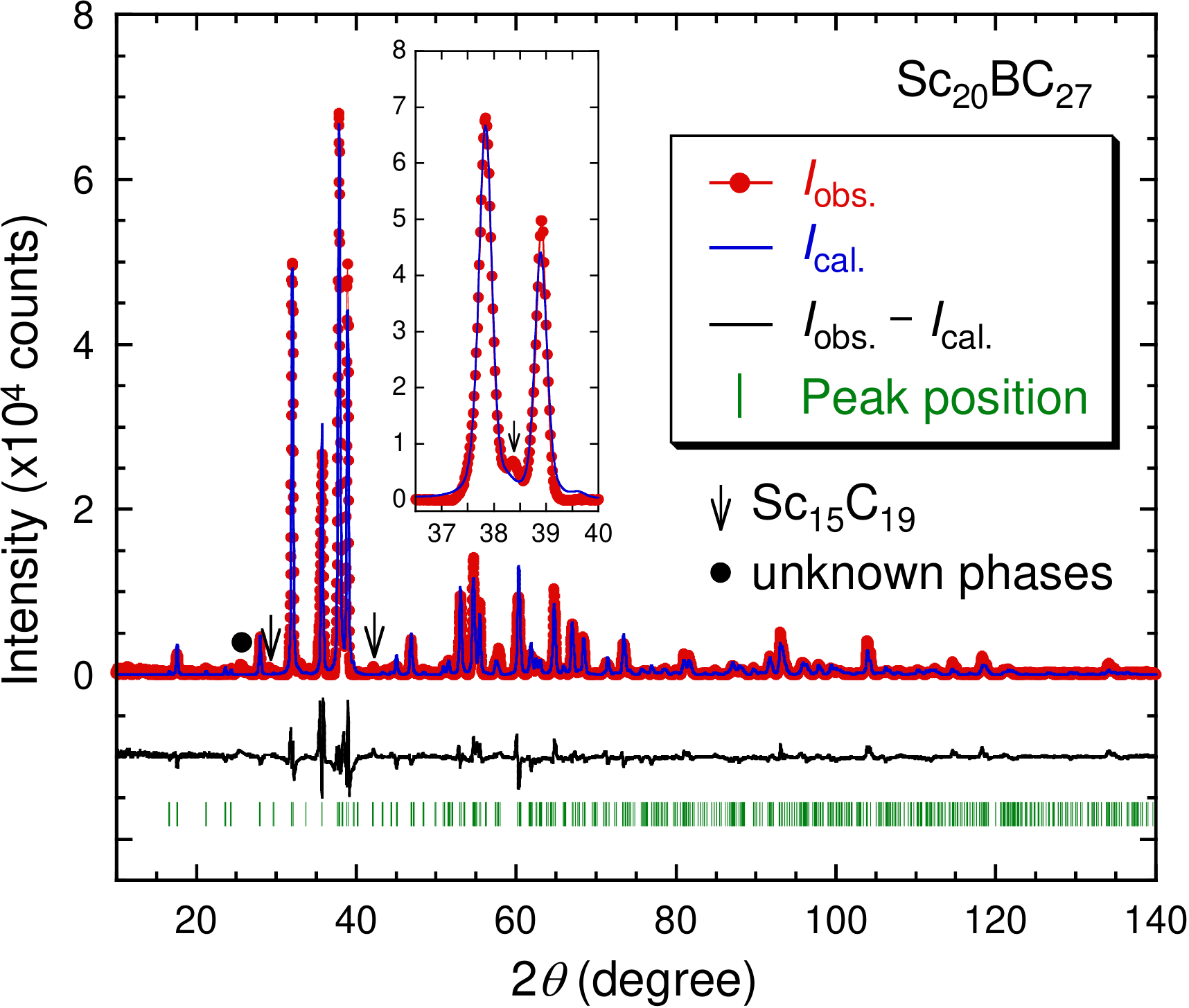}
	\caption{
	Powder X-ray diffraction pattern ($I_\mrm{obs.}$) and results from the Rietveld fitting ($I_\mrm{cal.}$) at approximately \SI{293}{K}. 
	The chemical composition was assumed to be \ce{Sc20BC27}.  
	A difference between $I_\mrm{obs.}$ and $I_\mrm{cal.}$ is shifted downward by $1 \times 10^{4}$\,counts. 
	Arrows and a circle indicate the peaks from \ce{Sc15C19} and unknown phases as impurities, respectively. 
	The inset depicts a magnification near the main peak. 
	}
	\label{XRD}
\end{figure}

We performed PXRD measurements using the sample that exhibited the largest diamagnetic signal, the nominal molar ratio of which was $\ce{Sc} : \ce{B} : \ce{C} = 35 : 2 : 63$. 
Figure \ref{XRD} illustrates the experimental PXRD pattern ($I_\mrm{obs.}$) and the results fromthe Rietveld fitting ($I_\mrm{cal.}$). 
In this refinement, peaks from a paramagnetic \ce{Sc15C19} and unknown phases as impurities were not in consideration because their intensity ratios of them to the main phase were sufficiently small. 
First, we performed structural optimization with the exception of B atoms, because distinguishing a small quantity of B from C atoms is difficult using the laboratory PXRD method. 
The observed diffraction peaks could be indexed from a tetragonal structure with a space group of $P4/ncc$ (\#130). 
This structural model, possessing the chemical formula of \ce{Sc20C28}, consisted of alternately stacked ScC- and C-layers along the crystallographic $c$-axis, the details of which are described below. 
Second, on the premise of this structure, we evaluated the enthalpy of formation using the DFT calculation to examine the atomic positions of B. 
Our calculations implied that the B atoms were the most stable at the C-layer labeled with the C3/B($8f$) site. 
For an isostructural ternary borocarbide of \ce{\tit{R}5B2C5}($R=\ce{Y},\, \ce{Ce}-\ce{Nd},\, \ce{Sm},\, \ce{Gd}-\ce{Tm}$), all the C atoms on the C-layer were replaced by B atoms\cite{R5B2C5_2000}, confirming the validity of the results of our DFT calculations. 
Given the optimized molar ratio for material synthesis, the quantity of B varied in the range of $x=0-2$ for \ce{Sc20C_{8-$x$}B_{$x$}C20}. 
Consequently, we finally refined the crystallographic parameters under the hypothesis that one of the C atoms at the $8f$ site randomly replaced a B atom in the unit cell; in other words, the chemical formula was assumed to be \ce{Sc20BC27}. 
To maintain the crystal symmetry of $P4/ncc$, the virtual chemical species of 0.875C+0.125B at the C3/B($8f$) site was practically employed because the PXRD pattern showed no extra peaks implying the formation of a superlattice structure by the periodic ordering of B atoms. 
The refined fractional coordinates are summarized in Table \ref{atom-position}. 
The best fit yielded the lattice constants of $a=\SI{7.5042(3)}{\angstrom}$ and $c=\SI{10.0386(5)}{\angstrom}$. 
The weighted-profile and expected reliability factors converged to $R_\mrm{wp}=0.138$ and $R_\mrm{e}=0.123$, respectively, resulting in a sufficiently low goodness-of-fit index value of $S=1.12$. 
Furthermore, both $R_\mrm{wp}$ and $R_\mrm{e}$ were not affected by the change in $x$ from 0 to 2 (hereafter, the 
chemical formula is referred to as \ce{Sc20BC27}).

\begin{table*}[!t]
	\caption{Atomic Positions, Site Occupancies ($Occ.$), and Isotropic Displacement Parameters ($U_\trm{iso}$) of a Superconducting \ce{Sc20BC27} Refined by a Rietveld Fitting at Approximately \SI{293}{K}. \textsuperscript{\emph{a}}}
	\label{atom-position}
\centering
\begin{tabular}[width=\linewidth]{lllllll}
\hline
\multicolumn{1}{c}{site} & \multicolumn{1}{c}{\begin{tabular}[c]{@{}c@{}}Wyckoff\\ letter\end{tabular}} & \multicolumn{1}{c}{\textit{x}} & \multicolumn{1}{c}{\textit{y}} & \multicolumn{1}{c}{\textit{z}} & \multicolumn{1}{c}{\textit{Occ.}} & \multicolumn{1}{c}{$U_\trm{iso}$} \\ \hline
    \multicolumn{1}{c}{Sc1} & \multicolumn{1}{c}{$16g$} & 0.0938(3) & 0.2132(3) & 0.0983(1) & \multicolumn{1}{c}{1} & 0.028(1) \\
    \multicolumn{1}{c}{Sc2} & \multicolumn{1}{c}{$4c$} & $1/2$ & 0 & 0.1361(4) & \multicolumn{1}{c}{1} & 0.028(1) \\
    \multicolumn{1}{c}{C1} & \multicolumn{1}{c}{$16g$} & 0.3948(16) & 0.2953(16) & 0.1324(5) & \multicolumn{1}{c}{1} & 0.028(1) \\
    \multicolumn{1}{c}{C2} & \multicolumn{1}{c}{$4c$} & 0 & $1/2$ & 0.1072(14) & \multicolumn{1}{c}{1} & 0.028(1) \\
    \multicolumn{1}{c}{C3/B}\textsuperscript{\emph{b}} & \multicolumn{1}{c}{$8f$} & 0.3292(15) & 0.3292(15) & $1/4$ & \multicolumn{1}{c}{1} & 0.028(1) \\ \hline
\end{tabular}

\textsuperscript{\emph{a}} Space group: $P4/ncc$(\#130, Origin choice 1), $Z=1$, $a=\SI{7.5042(3)}{\angstrom}$, $c=\SI{10.0386(5)}{\angstrom}$, $V=\SI{565.3}{\angstrom^3}$. 
The occupancy for each site was set to 1. 
Global isotropic displacement factors were employed for all sites. 
\textsuperscript{\emph{b}} The virtual chemical species of 0.875C+0.125B. 
\end{table*}

\begin{figure}[!t]
	\includegraphics[width=8.5cm,bb=0 0 659 530]{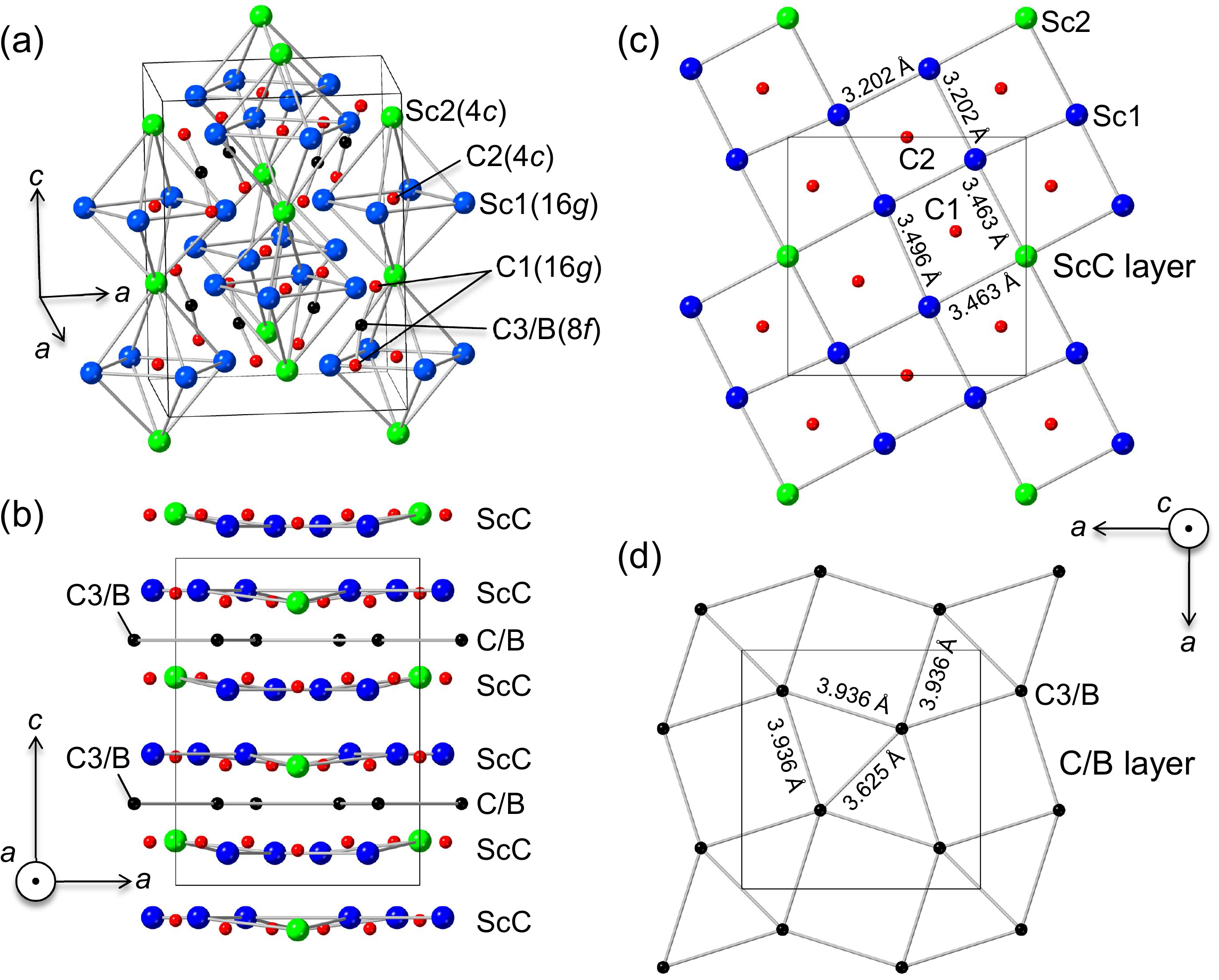}
	\caption{
	(a) Crystal structure of \ce{Sc20BC27} as determined by the Rietveld analysis. 
	(b) Layered structure along the crystallographic $c$-axis. 
	(c) ScC- and (d) C/B-layers in the $c$-plane. 
	The solid lines correspond to a unit cell. 
	}
	\label{structure}
\end{figure}

The crystal structure of \ce{Sc20BC27} is depicted in Figure \ref{structure}. 
As illustrated in Figure \ref{structure}a, every C2 atom was in an octahedral coordination with the Sc1 and Sc2 atoms. 
One of the heights from C2 to Sc2 atoms was \SI{2.577}{\angstrom}, and the other was \SI{2.442}{\angstrom}. 
The unit cell also included eight C1--C3/B--C1 units, whose angle was $165.2^\circ$. 
The interatomic distance of C1--C3/B--C1 units was calculated to be \SI{1.304}{\angstrom}, suggesting the double-bond character. 
A similar situation was observed in the isomorphic \ce{\tit{R}5B2C5}($R=\ce{Sm}\,\mrm{or}\,\ce{Gd}$)\cite{R5B2C5_2000}. 
If we consider as a two-dimensional layered structure composed of ScC- and C/B-layers, as shown in Figure \ref{structure}b, the stacking sequence of ScC--C/B--ScC along the $c$-axis was repeated twice in the unit cell. 
Each of the layers viewed from the $c$-axis is depicted in Figures \ref{structure}c and \ref{structure}d. 
In the ScC-layer, each C1 atom was located in a quasi-square lattice formed by the Sc1 and Sc2 atoms, whereas every C2 atom was surrounded by the Sc1 square sheets. 
The average Sc-Sc distance of approximately \SI{3.39}{\angstrom} was comparable to that of a pure Sc metal ($\sim \SI{3.31}{\angstrom}$)\cite{Sc-metal}. 
This layer was piled up along the $c$-axis direction rotated by $90^\circ$. 
The C/B layer, which was separated by two ScC-layers, consisted of the isosceles triangular and square nets. 
As described above, one or two of the eight C atoms were replaced with B atoms in this layer. 
The known typical B-poor or -free compounds, depicted in Figure \ref{contour}, crystallized in various layered structures that have an alternate stacking of \ce{ScC}- and C- or B-layers along the crystallographic $c$-axis\cite{Sc3C4-structure,Sc-B-C_system}, as in the case of \ce{Sc20BC27}. 
As for a series of \ce{\tit{R}5B2C5}, the C3/B($8f$) site was fully occupied by B atoms, leading to a larger unit-cell volume\cite{R5B2C5_2000}. 
This may be because the ionic radii of the other rare-earth metals are relatively larger than that of Sc.

\subsection{Superconducting Properties}

\begin{figure}[!t]
	\includegraphics[width=8cm,bb=0 0 434 419]{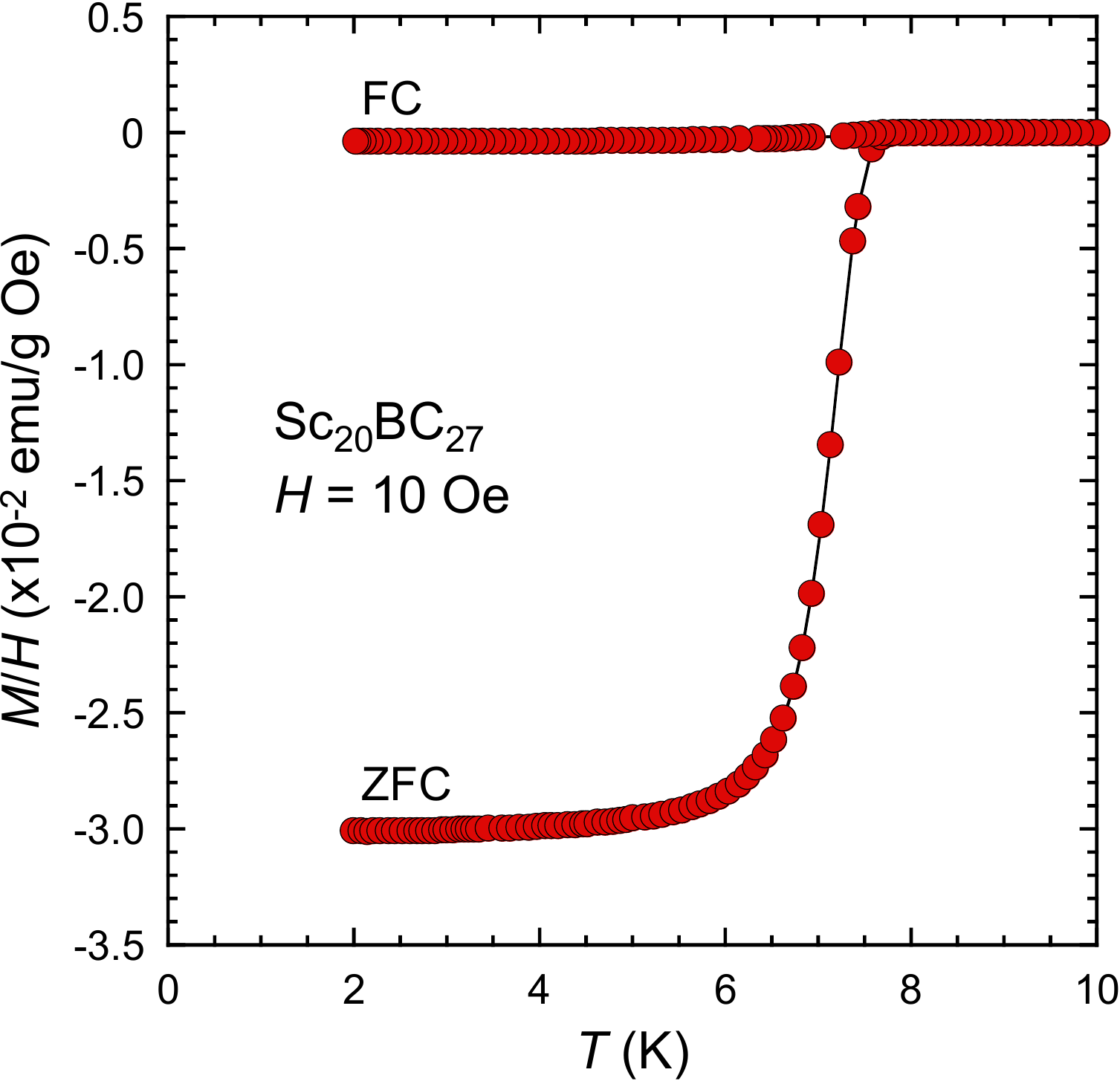}
	\caption{Temperature dependence of magnetic susceptibility ($M/H$) measured under a field of \SI{10}{Oe}. 
	ZFC and FC indicate the zero-field-cooling and field-cooling processes, respectively. }
	\label{MT10Oe}
\end{figure}

The magnetic susceptibility $(M/H)$ of \ce{Sc20BC27} is plotted as a function of temperature in Figure \ref{MT10Oe}. 
We observed a clear diamagnetic signal related to the superconducting transition at the onset of $T_\trm{c}=\SI{7.7}{K}$. 
Because the shielding-volume fraction at \SI{2}{K} exceeded \SI{130}{\%} without correcting the demagnetization factor, it is sufficient to regard the sample as a bulk superconductor. 

\begin{figure}[!t]
	\includegraphics[width=8cm,bb=0 0 286 414]{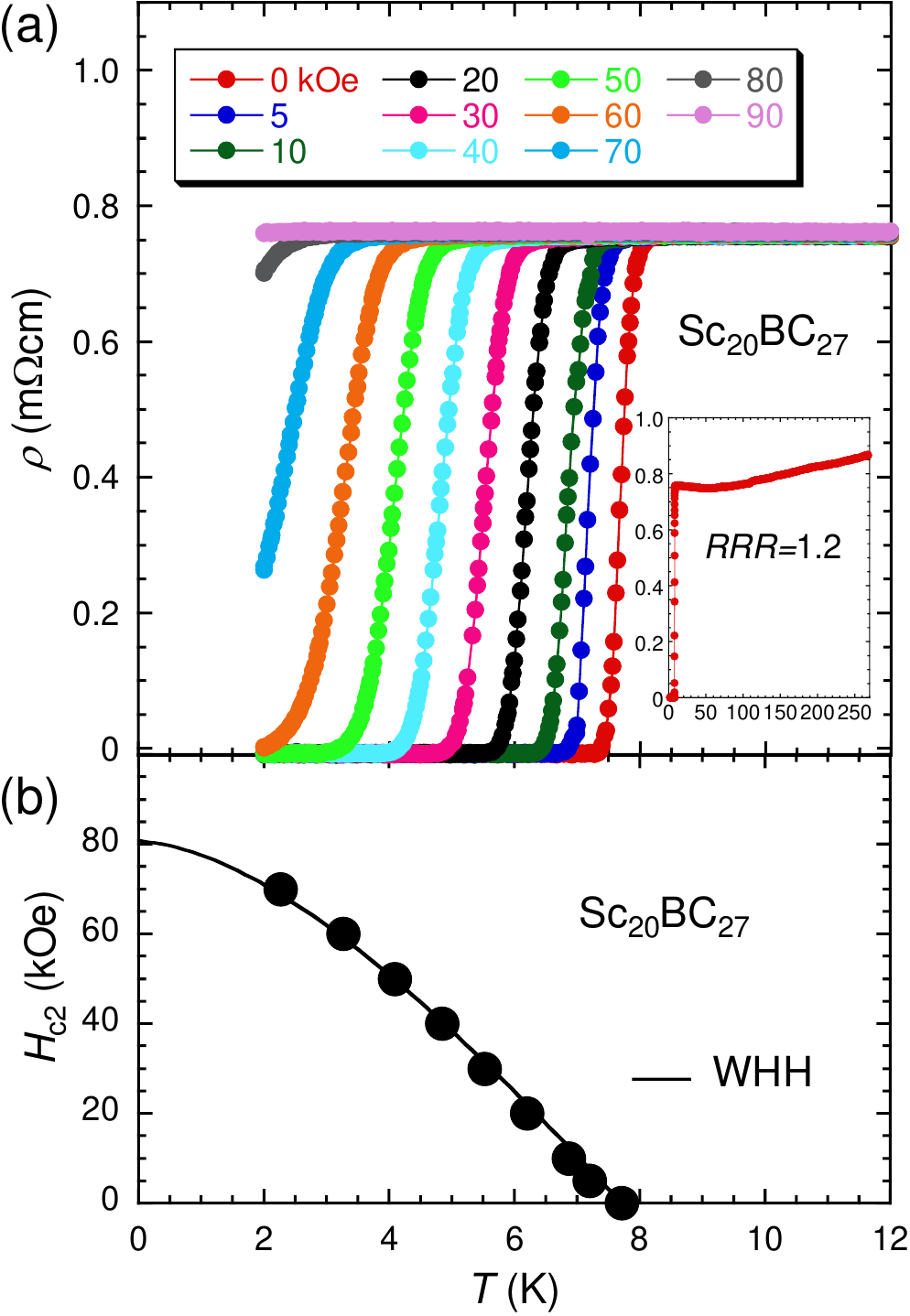}
	\caption{(a) Electrical resistivity $(\rho)$ as a function of temperature for \ce{Sc20BC27} measured under various magnetic fields up to \SI{90}{kOe}. 
	The inset depicts the zero-field $\rho - T$ curve below approximately room temperature. 
	$RRR$ represents the residual resistivity ratio (=$\rho_\mrm{270\,K} / \rho_{0}$), where $\rho_0$ is the $\rho$ value at the onset of the resistive transition.  
	(b) Temperature dependence of the upper critical field $H_\mrm{c2}$. 
	The solid curve represents the calculated result on the basis of the WHH theory \cite{WHH,WHH-fitting}.
	}
	\label{rho_Hc2}
\end{figure}

Figure \ref{rho_Hc2}a illustrates the temperature dependences of electrical resistivity $(\rho)$ measured at several fields up to \SI{90}{kOe}. 
The inset depicts the $\rho - T$ curve below room temperature at the zero field. 
With decreasing temperature, the resistivity exhibited a normal metallic behavior and subsequently showed a slight upturn at around \SI{50}{K}. 
Below $T_\mrm{c}^\mrm{onset}=\SI{8.3}{K}$, an abrupt decrease arising from the superconducting transition was observed. 
As a result, zero resistivity was attained at $T_\mrm{c}^\mrm{zero}=\SI{7.7}{K}$. 
This value of $T_\mrm{c}^\mrm{zero}$ corresponds to the onset temperature observed in the susceptibility. 
The residual resistivity ratio $(RRR)$ was estimated to be a maximum of 1.2, as the resistivity hardly depends on the temperature in the normal state. 
A similar feature was reported on the resistivity of some carbides \cite{Mo3Al2C_2010PRB} and B-doped diamond\cite{B-doped_diamond_Tc=4K}, suggesting that the transport properties on a polycrystalline specimen are sensitive to residual C in the grain boundaries\cite{W7Re13X_Kawashima}. 
This is probably why the slight upturn of the $\rho-T$ curve was observed. 
As illustrated in Figure \ref{rho_Hc2}a, with increasing field, we observed a parallel shift of the $T_\mrm{c}$ to a lower temperature, and no trace of superconductivity was observed down to \SI{2}{K} at $H=\SI{90}{kOe}$. 
Figure \ref{rho_Hc2}b illustrates the upper critical field $H_\mrm{c2}$ as a function of temperature. 
We defined the $T_\mrm{c}$ from the midpoint of the superconducting transition. 
As indicated by the solid curve, $H_\mrm{c2}(T)$ obeyed the Werthamer-Helfand-Hohenberg (WHH) prediction\cite{WHH,WHH-fitting}, resulting in a value of $H_\mrm{c2}(0)=\SI{80.5}{kOe}$. 
We calculated the Ginzburg-Landau (GL) coherence length of $\xi_0 = \SI{6.4}{nm}$ using the relation of $H_\mrm{c2}(0)=\phi_0/2\pi\xi_0^2$ ,where $\phi_0$ is the magnetic flux quantum. 

\begin{figure}[!t]
	\includegraphics[width=8cm,bb=0 0 279 528]{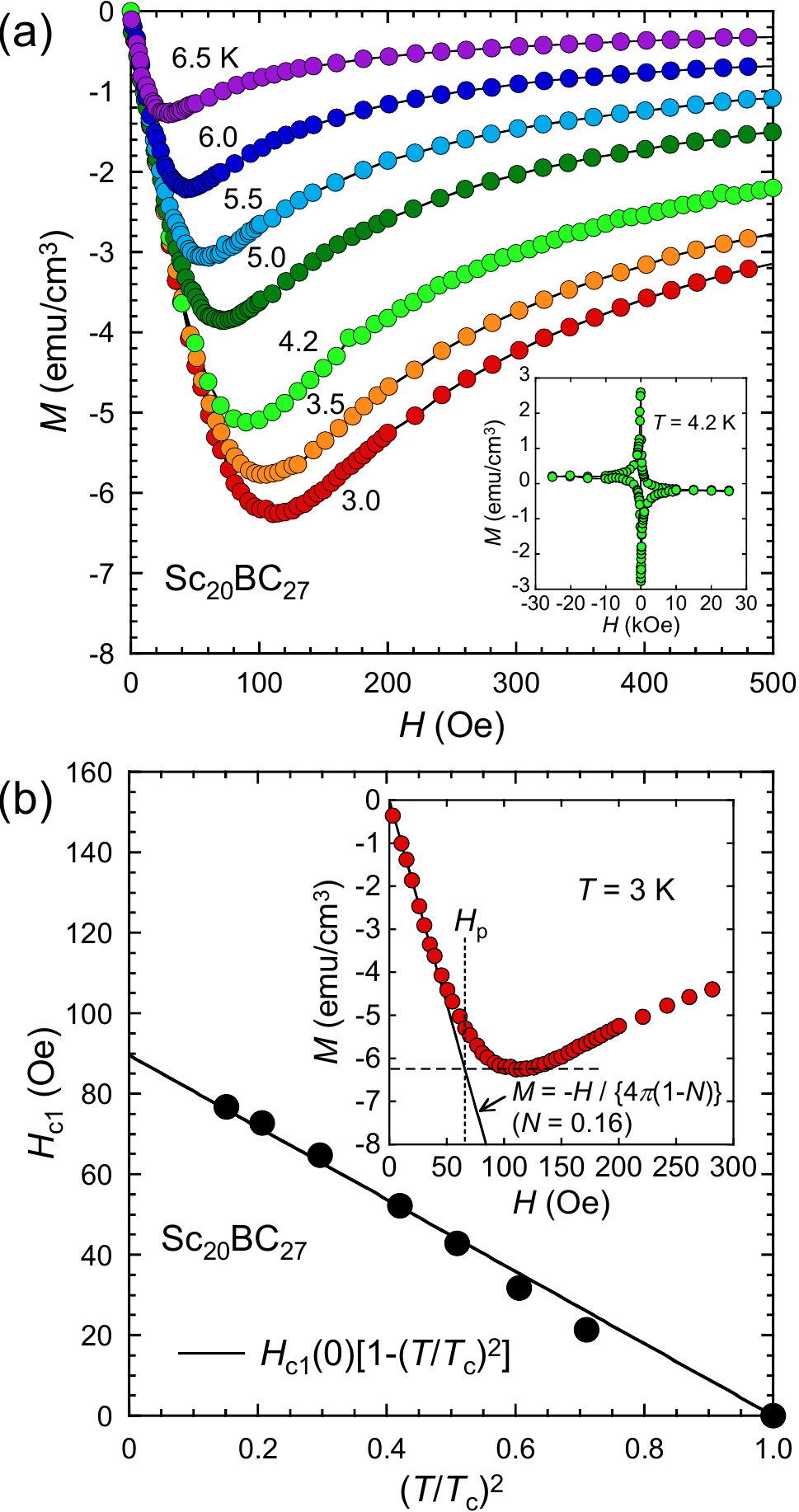}
	\caption{(a) Magnetic-field dependence of the magnetization for \ce{Sc20BC27} measured at several temperatures down to \SI{3.0}{K}. 
	The inset depicts a magnetization-hysteresis loop at \SI{4.2}{K}. 
	(b) Relation between lower critical field $H_\mrm{c1}$ and squared normalized temperature $(T/T_\trm{c})^2$. 
	The solid line indicates a fitting result by the formula. 
	The inset depicts an enlarged view of the magnetization process below \SI{300}{Oe} measured at $T=\SI{3}{K}$. }
	\label{MH}
\end{figure}

Figure \ref{MH}a illustrates the magnetization as a function of the magnetic field measured at various temperatures for \ce{Sc20BC27}. 
Additionally, a magnetization-hysteresis loop at $T=\SI{4.2}{K}$ is depicted in the inset of Figure \ref{MH}a. 
This magnetization behavior is typical of type-II superconductors. 
At $T=\SI{3}{K}$, the magnetization exhibited a broad minimum at approximately \SI{100}{Oe} and then increased gradually up to \SI{500}{Oe}. 
The minimum of the $M-H$ curve was suppressed, which then shifted to the lower field as the temperature approached the $T_\mrm{c}$. 
On the basis of these data, we evaluated the lower critical field $H_\mrm{c1}$, which is plotted as a function of the squared normalized temperature $(T/T_\mrm{c})^2$ in Figure \ref{MH}b. 
An enlarged version of the $M-H$ curve at \SI{3}{K} is depicted in the inset of Figure \ref{MH}b. 
In this study, $H_\mrm{c1}$ was approximated from a relation of $H_\mrm{c1}=H_\mrm{p}/(1-N)$\cite{Hc1-demag}, where $H_\mrm{p}$ and $N$ are the penetration field and demagnetization factor\cite{DemagnetizingFactor}, respectively. 
$H_\mrm{p}$ was defined as the intersection point between the extrapolation of the perfect diamagnetism $(M=-H/{4\pi(1-N)})$ and a level of the minimum magnetization (dashed line)\cite{Hc1-demag}. 
We estimated $N$ to be 0.16 from the initial slope of the $M-H$ curve. 
Note that the values of $H_\mrm{c1}$ were possibly overestimated in this definition. \cite{BaNi2P2_PRB2009}
As indicated by the solid line in Figure \ref{MH}b, it was found that $H_\mrm{c1} - (T/T_\mrm{c})^2$ data was well reproduced by the formula based on the GL theory of $H_\mrm{c1}(T)=H_\mrm{c1}(0)[1-(T/T_\mrm{c})^2]$. 
We estimated $H_\mrm{c1}(0)$ to be \SI{90}{Oe} and calculated the London penetration depth of $\lambda_0 \approx \SI{271}{nm}$ using an approximation of $H_\mrm{c1}(0) \approx \phi_0 / \pi \lambda_0^2$. 
The GL parameter of $\kappa_\mrm{GL} = \lambda_0 / \xi_0$ was also determined to be $\kappa_\mrm{GL} \approx 42 (> 1/\sqrt{2})$, strongly supporting the type-II superconducting nature. 

\begin{figure}[!t]
	\includegraphics[width=8cm,bb=0 0 284 530]{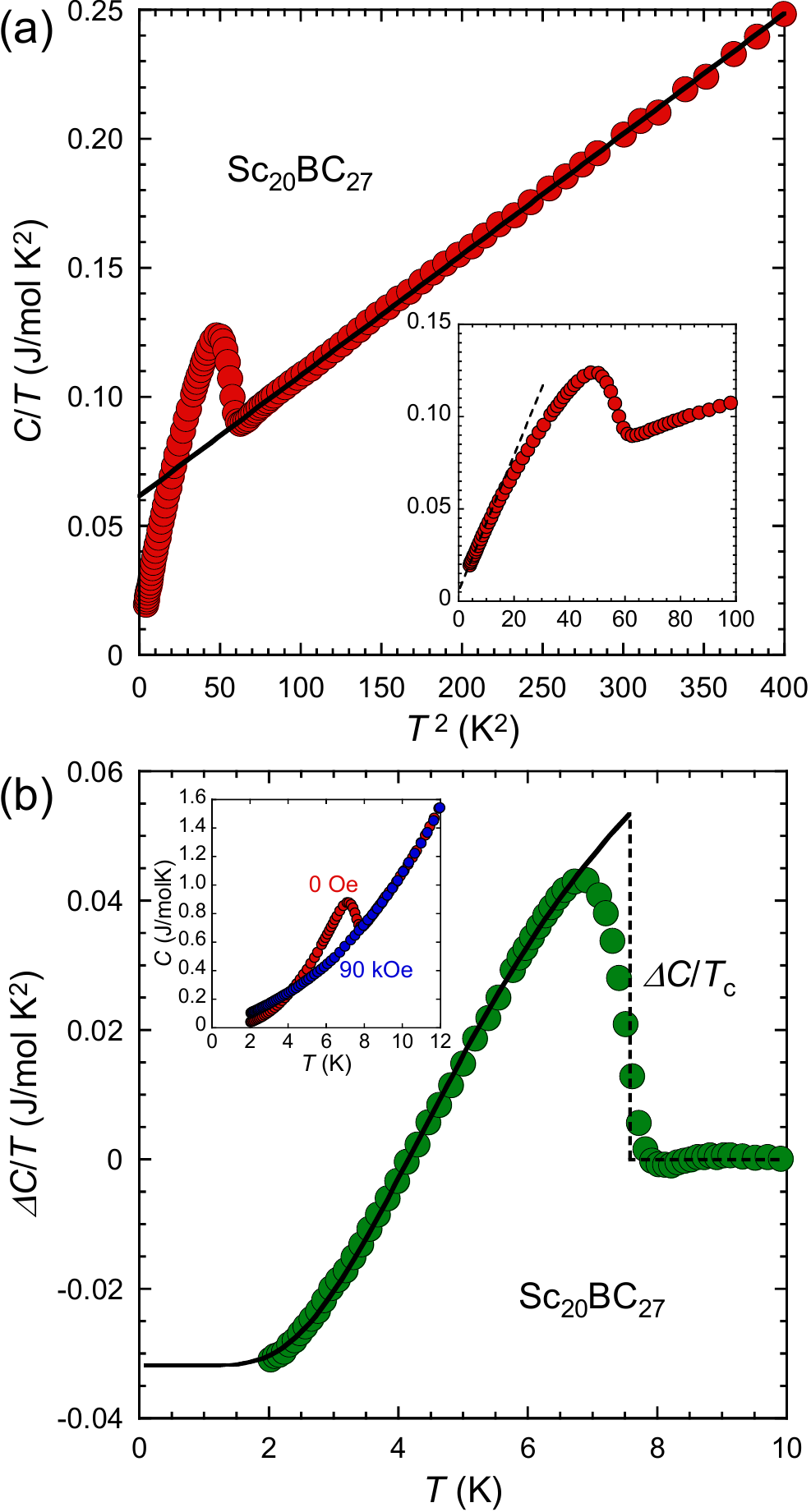}
	\caption{
	(a) Specific heat divided by temperature $(C/T)$ as a function of the squared temperature for \ce{Sc20BC27} measured at the zero field. 
	The solid line represents the linear extrapolation from the normal state to \SI{0}{K}. 
	The inset depicts an enlarged view of the $C/T-T^2$ curve below \SI{10}{K}. 
	The dashed line is a linear fit to the data below \SI{3}{K}. 
	(b) Temperature dependence of a specific-heat difference between the superconducting (\SI{0}{kOe}) and normal (\SI{90}{kOe}) states. 
	The solid line represents a calculated $\Delta C/T$ curve assuming an isotropic single gap. 
	The inset depicts the $C-T$ curves measured at the zero field and \SI{90}{kOe}. 
	}
	\label{CT}
\end{figure}

Figure \ref{CT}a illustrates the dependences on the squared temperature of the specific heat divided by temperature $(C/T)$ for \ce{Sc20BC27}. 
The specific heat demonstrated a sharp increase below \SI{7.7}{K}($T^2=\SI{60}{K^2}$). 
This temperature is in good agreement with the $T_\mrm{c}$ examined by the susceptibility and resistivity measurements, clarifying the bulk nature of superconductivity. 
At the normal state from the $T_\mrm{c}$ to \SI{20}{K}, the data followed the sum of the electronic and phononic parts generally expressed by $C/T = \gamma_\mrm{n} + \beta T^2$, where $\gamma_\mrm{n}$ and $\beta$ denote the Sommerfeld coefficient at the normal state and the phonon contribution to the specific heat, respectively. 
As indicated by the solid line, the linear extrapolation to \SI{0}{K} yielded $\gamma_\mrm{n} = \SI{61.5}{mJ/mol K^2}$, $\beta = \SI{0.468}{mJ/molK^4}$, and the corresponding Debye temperature of $\Theta_\mrm{D}=\SI{584}{K}$, calculated using $\beta = 12 \pi^4 NR/5 \Theta_\mrm{D}^3$, where $N$ is the number of atoms in a formula and $R$ is the gas constant. 
However, the sample of \ce{Sc20BC27} contained non-negligible amounts of \ce{Sc15C19} and unassigned nonsuperconducting impurities, which could be observed in the PXRD pattern. 
As indicated by the dashed line in the inset of Figure \ref{CT}a, a residual Sommerfeld coefficient, resulting from the impurities, was estimated to be $\gamma_\mrm{r} = \SI{7.0}{mJ/molK^2}$ by extrapolating the data to $T=\SI{0}{K}$. 
To exclude the influence of the secondary phases, using raw data, we subtracted the $C-T$ data measured at \SI{90}{kOe} from that in the zero field. 
The difference $C$ between the superconducting and normal states ($\Delta C$) is plotted as $\Delta C/T - T$ in Figure \ref{CT}b. 
As shown in the inset, the $C-T$ curve at \SI{90}{kOe} exhibited no anomalies, indicating that the superconducting transition was completely suppressed below \SI{2}{K}. 
Assuming the magnitude of a single superconducting gap $\varDelta_0$, the specific heat below $T_\mrm{c}$ can be expressed as $\Delta C/T = \gamma_\mrm{s} + A \exp(-\varDelta_0/T)/T$, where $\gamma_\mrm{s}$ corresponds to an electronic specific-heat coefficient of the superconducting phase and $A$ is a free parameter. 
As indicated by the solid curve, the fit to the data gave $\gamma_\mrm{s}=\SI{31.8}{mJ/molK^2}$ and $\varDelta_0 = \SI{1.26}{meV}$. 
Both values of $\gamma_\mrm{n}$ and $\gamma_\mrm{s}$ agree in order of magnitude with that evaluated by the first-principles calculation, which is discussed in the next section. 
The good reproducibility of $\Delta C/T-T$ proposes the single-gap \tit{s}-wave superconductivity in \ce{Sc20BC27}. 
Additionally, we estimated the normalized specific-heat jump at $T_\mrm{c}$ of $\Delta C/T_\mrm{c}=\SI{0.053}{J/molK^2}$, and calculated $\Delta C/\gamma_\mrm{s} T_\mrm{c}$ to be 1.67, which is slightly larger than the BCS weak-coupling limit of 1.43. 
For conventional superconductors, an empirical electron-phonon coupling constant $\lambda_\trm{e-p}$ is given by
\begin{equation*} 
\lambda_\trm{e-p}=\frac{1.04+\mu^{*} \ln \left(\varTheta_\mrm{D} / 1.45 T_\mrm{c}\right)}{\left(1-0.62 \mu^{*}\right) \ln \left(\varTheta_\mrm{D} / 1.45 T_\mrm{c}\right)-1.04}
\end{equation*}
where $\mu^\ast = 0.13$ is the Coulomb pseudopotential for polyvalent transition metals\cite{Macmillan}. 
Using this formula, $\lambda_\trm{e-p}$ was evaluated to be 0.60, thereby revealing that \ce{Sc20BC27} was an $s$-wave superconductor in the intermediate coupling regime. 

\subsection{Electronic-Band-Structure calculations}

\begin{figure}[!t]
	\includegraphics[width=8cm,bb=0 0 338 488]{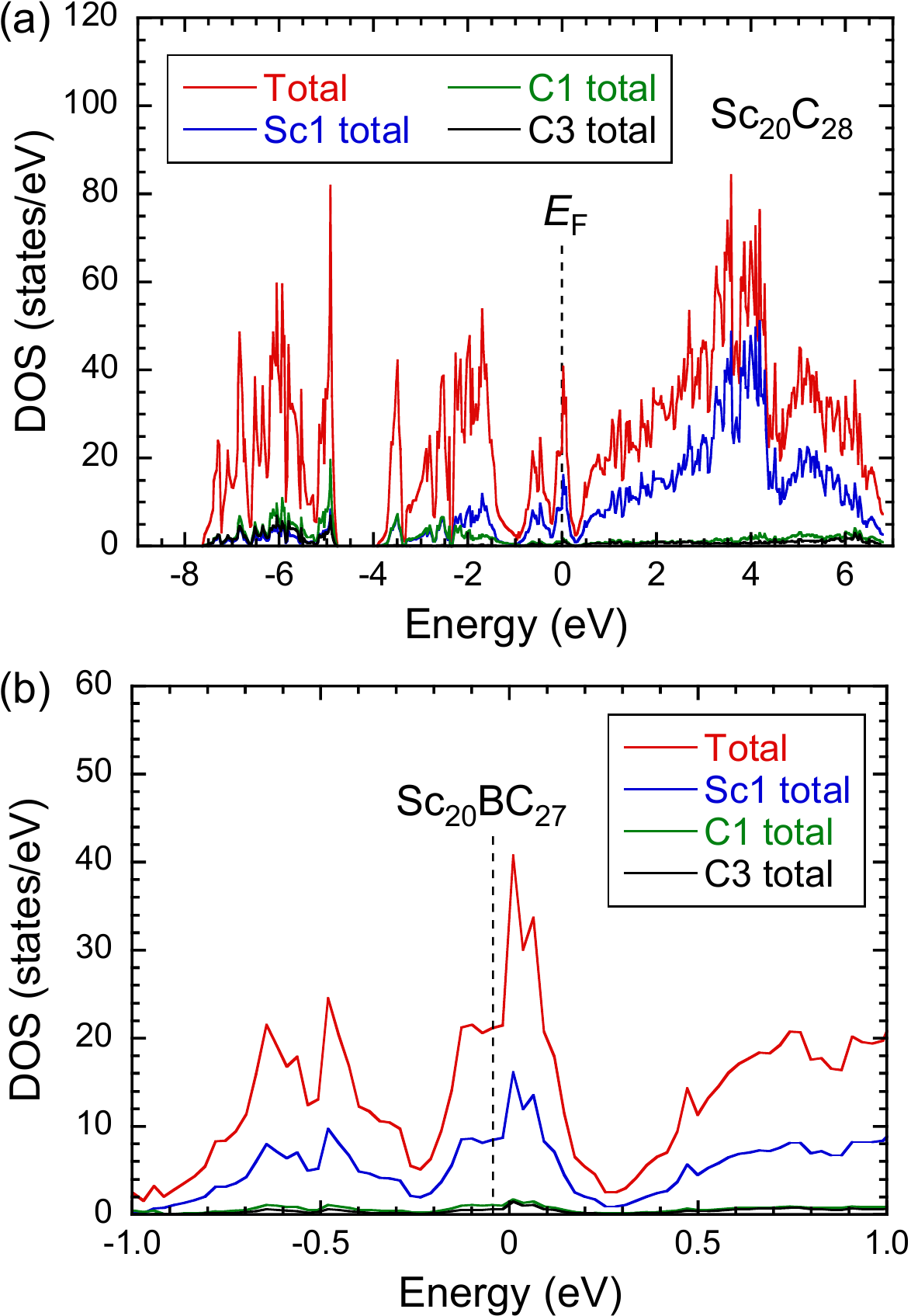}
	\caption{
	(a) Total and partial density of states (DOS) curves for \ce{Sc20C28} elucidated by a first-principles calculation. 
	$E=\SI{0}{eV}$ corresponds to the Fermi energy of $E_\trm{F}$. 
	(b) Magnified DOS curves in the vicinity of $E_\trm{F}$. 
	The $E_\trm{F}$ for \ce{Sc20BC27}, which is determined assuming the rigid-band model, is indicated by the vertical dashed line. 
	}
	\label{DOS}
\end{figure}

First, we calculated the DOS of \ce{Sc20C28} using the experimentally determined crystallographic parameters, as presented in Table \ref{atom-position}. 
Next, to discuss the electronic structure of \ce{Sc20BC27}, we applied a rigid-band model by shifting the Fermi energy ($E_\trm{F}$). 
The obtained DOS curve of \ce{Sc20C28} is illustrated in Figure \ref{DOS}a. 
A peak structure was observed at approximately $E_\trm{F}$, which was mainly constructed by the DOS derived from the Sc $3d$ orbitals. 
The value of $D(E_\trm{F})$ was calculated to be \SI{34.4}{eV^{-1}} per formula unit, resulting in a Sommerfeld coefficient of $\gamma_{\ce{Sc20C28}}=\SI{81.2}{mJ/molK^2}$. 
Figure \ref{DOS}b illustrates the enlarged DOS curve in the vicinity of $E_\trm{F}$. 
The contributions of the C1 and C3 atoms to the total DOS were negligibly small near $E_\mrm{F}$. 
This situation enabled us to adopt the rigid-band model\cite{Iyo-SrBi3_SciRep,Iyo-SrGe2,Hase-SrGe2}.  
When one of C atoms replaced a B atom in the unit cell, $E_\trm{F}$ shifted by $\sim \SI{0.045}{eV}$, as indicated by the dashed line in Figure \ref{DOS}b. 
On the basis of this, $D(E_\trm{F})$ and the corresponding $\gamma_{\ce{Sc20BC27}}$ were estimated to be \SI{21.2}{eV^{-1}}/f.u. and \SI{50.4}{mJ/molK^2}, respectively.  
The calculated $\gamma$ value was comparable with that obtained by the specific-heat measurements. 
Among the isostructural family of \ce{\tit{R}5B2C5}, the $R=\ce{Y}$ member was a Pauli paramagnet\cite{R5B2C5_2000}, although both Sc and Y are isovalent nonmagnetic metals. 
This suggests that the smaller ionic radius and lighter mass of Sc play important roles in the emergence of superconductivity, which supports the main contribution of Sc to the total DOS at $E_\mrm{F}$. 

By accepting the results of the \tit{ab initio} calculations, we can expect to realize the higher $T_\trm{c}$ in B-free \ce{Sc20C28} because the $D(E_\trm{F})$ value is larger than that of \ce{Sc20BC27}. 
However, as described above, the present crystal structure needs to be stabilized with a small quantity of B. 
One possible reason for this is that, as can be observed from Figure \ref{DOS}b, the $E_\mrm{F}$ of \ce{Sc20C28} is closer to the DOS peak, which implies the relative phase instability of \ce{Sc20C28} in comparison with \ce{Sc20BC27}. 
On the other hand, we could not synthesize the B-excess samples. 
With the increase in content of B, the $E_\trm{F}$ shifted to the lower-energy region. 
As depicted in Figure \ref{DOS}b, although the DOS curve was almost constant at approximately $E_\trm{F}$ of \ce{Sc20BC27}, it demonstrated a rapid decrease below \SI{-0.12}{eV}. 
Hence, even if heavily B-doped phases can be obtained, they will exhibit a lower $T_\mrm{c}$ or no superconductivity owing to the small DOS. 
In other words, if electron-doped phases are synthesized using other possible methods, the $T_\mrm{c}$ will be slightly increased. 

\section{Conclusion}

The polycrystalline samples of ternary borocarbide \ce{Sc20C_{8-$x$}B_{$x$}C20} $(x=1\ \trm{or}\ 2)$ were successfully synthesized. 
The Rietveld refinements demonstrated that the crystal structure belongs to the tetragonal space group of $P4/ncc$, and possesses the distorted \ce{Sc6} octahedra and eight units formed by the \ce{C3} group. 
The unit cell was also regarded as a layered structure made of ScC- and C-meshes along the $c$-axis.  
Using the DFT calculations to examine the positional stability of B, we found that one or two C atoms at the $8f$ site (isolated C-layer) were replaced by B. 
The magnetic susceptibility, electrical resistivity, and specific heat data revealed that the compound exhibited the typical type-II superconductivity at $T_\mrm{c}=\SI{7.7}{K}$.  
To summarize, the physical parameters determined in this study are listed in Table \ref{parameter}. 
The magnitude of the specific-heat jump and the electron-phonon coupling constant elucidated that the superconductivity in \ce{Sc20C_{8-$x$}B_{$x$}C20} could be classified as an intermediately coupled system. 
The DOS curve indicated that the Sc-$3d$ orbitals principally contributed to the total DOS at $E_\mrm{F}$. 
Additionally, using the rigid-band model, we provided the qualitative explanation for the result that the structure containing a small quantity of B was rather stable; however, the higher $T_\mrm{c}$ was expected in the B-free composition of $x=0$. 

\begin{table}[!t]
\centering
\caption{Physical Parameters at the Normal and Superconducting States for \ce{Sc20BC27}. }
\label{parameter}
\begin{tabular}{cc}
\hline
parameters & \ce{Sc20BC27} \\ \hline
$T_\mrm{c}$& \SI{7.7}{K}\\
$H_\mrm{c1}(0)$& \SI{90}{Oe} \\
$H_\mrm{c2}(0)$& \SI{80.5}{kOe} \\
$\lambda_0$& \SI{271}{nm} \\
$\xi_0$& \SI{6.4}{nm} \\
$\kappa_\mrm{GL}$& 42 \\
$\gamma_\mrm{s}$& \SI{31.8}{mJ/molK^2}\\
$\beta$& \SI{0.468}{mJ/molK^4} \\
$\varTheta_\mrm{D}$& \SI{584}{K} \\
$\Delta C/\gamma_\mrm{s} T_\mrm{c}$& 1.67 \\
$\lambda_\trm{e-p}$ & 0.60 \\ \hline
\end{tabular}
\end{table}

\begin{acknowledgement}
 
The authors thank Editage [\url{http://www.editage.com}] for English language editing and reviewing of this manuscript. 

\end{acknowledgement}






\providecommand{\latin}[1]{#1}
\makeatletter
\providecommand{\doi}
  {\begingroup\let\do\@makeother\dospecials
  \catcode`\{=1 \catcode`\}=2 \doi@aux}
\providecommand{\doi@aux}[1]{\endgroup\texttt{#1}}
\makeatother
\providecommand*\mcitethebibliography{\thebibliography}
\csname @ifundefined\endcsname{endmcitethebibliography}
  {\let\endmcitethebibliography\endthebibliography}{}

\end{document}